# Friction and Memory Effects in Homogeneous Gas-Liquid Nucleation: Quest for Quantitative Rate Calculation

**Subhajit Acharya and Biman Bagchi\***

Solid State and Structural Chemistry Unit

Indian Institute of Science, Bengaluru, India

*corresponding author's email: profbiman@gmail.com, bbagchi@iisc.ac.in

## Abstract

The task of a first principles theoretical calculation of the rate of gas-liquid nucleation has remained largely incomplete despite the existence of reliable results from unbiased simulation studies at large supersaturation. Although the classical nucleation theory (CNT), formulated by Becker-Doring-Zeldovich (BDZ) about a century ago, provides an elegant, widely-used picture of nucleation in a first-order phase transition, the theory finds difficulties in *predicting the rate accurately*, especially in the case of gas-to-liquid nucleation. Here, we use a multiple-order parameter description to construct the nucleation free energy surface needed to calculate the nucleation rate. A multidimensional non-Markovian (MDNM) rate theory formulation that generalizes Langer's well-known nucleation theory by using Grote-Hynes multidimensional non-Markovian treatment is used to obtain the rate of barrier crossing. *We find good agreement of the theory with the rate obtained by direct unbiased molecular dynamics simulations—the latter is feasible at large supersaturation,* S. The theory gives the experimentally strong dependence of the rate of nucleation on supersaturation, S. Interestingly, we find a strong influence of frequency-dependent friction at the barrier top. This arises from multiple recrossing of the barrier surface. We find that a Markovian theory, such as Langer's formulation, fails to capture the rate quantitatively. In addition, the multidimensional transition state theory expression performs poorly, revealing the underlying role of friction.

# I. INTRODUCTION

Phase transitions are omnipresent in nature----from the formation of raindrops in atmospheric science to the freezing and melting of liquids and solids, magnetic transitions, as well as to understand crystallization of polymers, proteins, and helix-coil transitions of DNA.[1–5] In the realm of everyday experience, the majority of observed phase transitions[6] exhibit a first-order nature, and the kinetics of these transitions involve two pivotal stages, namely, nucleation and growth, with nucleation as the initial step in forming the liquid phase. Nucleation is usually categorized into two classes: homogeneous and heterogeneous, with the former being the subject of present research.[7]

In homogeneous gas-to-liquid nucleation, gas-phase molecules spontaneously coalesce to form stable nuclei without external influence or the presence of foreign particles. This process typically occurs in a supersaturated gas or vapor, as the gas becomes thermodynamically unstable and begins forming liquid droplets or bubbles. Homogeneous nucleation is characterized by a nucleation rate that is dependent on factors like the degree of supersaturation, temperature, pressure, and molecular interactions. Once a nucleus becomes larger than a critical size, it can rapidly grow through the condensation of gas molecules onto the liquid surface. The formation of the critical nucleus is an activated process. At low supersaturation or low supercooling, the activation barrier is large on the scale of the thermal energy, $k_BT$, where $k_B$ is the Boltzmann constant, and $T$ is the temperature, and the process is slow, especially at low supersaturation, $S$.[1]

Over the last few decades, gas-liquid nucleation has been extensively studied through experiments, simulations, and theory.[8–10] The rare-event nature of critical nucleus formation, which may take seconds or longer, places it well beyond the MD simulation time scale. Direct

MD simulations can only capture nucleation events at high supersaturations[11] or in expensive, massively parallel large-scale calculations.[12] Additionally, the use of computationally efficient small simulation cells introduces significant finite-size artifacts.[13] In this context, one needs to perform enhanced sampling techniques like metadynamics or umbrella sampling to study nucleation at low supersaturation. Recent computational studies indeed have employed umbrella sampling and metadynamics to map out the reaction-free energy surface in terms of several collective variables, referred to as order parameters.[14,15] We choose two-order parameters, as seen in a recent study on insulin dimer dissociation.[16] However, one often faces multiple choices of order parameters, with the decision guided by the questions we pose and the answers we seek.[17]

In the simple Becker-Doring-Zeldovich (BDZ) classical nucleation theory, one employs the radius of the nucleus (assumed to be spherical) as the order parameter, which is then used under capillary approximation to obtain the free energy barrier involved in the formation of a critical nucleus.[18,19] Among multiple approximations involved in the BDZ approach, the use of capillary approximation poses a serious limitation. This approach also ignores multiple fluctuations like the shape and the state of the surrounding gas particles. In the density functional theory developed by Oxtoby and Talaquar, one employs a position-dependent density profile as the order parameter; the free energy is minimized with respect to this density profile.[20] This quasi-analytical approach requires the input of the two and three-particle distribution functions, which are hard to obtain in situations where the density varies sharply across the profile.

The difficulties of obtaining an accurate nucleation free energy surface stem from the issues related to the many-body aspect or the collective nature of the problem. *We need to find the free energy of the entire system that contains clusters or nuclei of all sizes*. We need to identify the largest liquid-like cluster that can attain the critical size.

We now discuss the selection of order parameters. We start by mentioning two relevant studies here. In the study of Bhimalapuram *et al.*, the two order parameters chosen were the density of the liquid and the liquidness of the system. The latter was defined as the number ($n_{liq}$) of liquid-like particles (LLP).[21] This was a quantity introduced by Stillinger. An LLP is a particle that contains more or equal to five neighbors. In a different work, Tiwari and coworkers employed different order parameters to describe the emergence of a liquid-like cluster, namely, (i) $n_{liq}$ and (ii) local density fluctuations. The latter allows for shape fluctuation and captures some aspects of the size of the cluster.[22] We have selected the same order parameters employed by Tiwari and coworkers. They are (i) the liquidness of the system ($n_{liq}$) and (ii) fluctuations in the coordination number of a cluster ($\mu_2$). It is essential to realize that these physically motivated sets of order parameters are not fully orthogonal, but they do contain a portion of orthogonal information that can be extracted. Secondly, they are not exhaustive. The third-order parameter that we monitor is the size of the largest cluster, denoted by $n_{largest}$. At any given time, one can calculate the distribution of cluster sizes and determine the largest cluster. What makes this parameter unique is that one always finds that as the nucleation point is approached, only one large cluster survives that grows to become the critical size. As there is a large separation between the largest cluster and the remaining cluster sizes, there is no ambiguity in following the largest cluster, as shown later in **Figure 4**.

The idea of a liquid-like cluster was introduced by Stillinger, who defined a liquid-like particle as one that has at least five nearest neighbors.[23] A liquid-like cluster is a cluster made of liquid-like particles. Thus, a monomer of a liquid-like cluster has at least five nearest-neighbor particles and can have more. Therefore, a liquid-like cluster differs from an ordinary cluster, where a cluster of size *n* contains just *n* particles. However, we can have a relatively large simple cluster of *n* particles, which is so ramified that it does not contain a single liquid-like particle. A liquid-like cluster can represent a liquid droplet. As the size of a liquid-like cluster grows, its size (in terms of the number of liquid-like particles it contains) approaches the size given by the sheer number of particles. This is an important attribute.

The second-order parameter describes the fluctuation in coordination number, which represents the compactness of a cluster. When the cluster size increases, the growth of fluctuation slows down, and for a liquid-like cluster beyond the critical size, this fluctuation sharply decreases.

Once we have determined the free energy surface as a function of the order parameters, the aim is to calculate the rate of nucleation, which is the main emphasis of this article. In contrast to the focus placed on determining the multidimensional reaction-free energy surface, less effort has been devoted to the calculation of *the rate by quasi-analytical means*. The most notable work here is the seminal study of Langer, whose work has shaped much of our discussions on nucleation.[24] Such a calculation requires the details of the free energy surface, similar to the approach used for years with the one-dimensional reaction surface. In fact, even though the study of Langer remains a landmark in this field, detailed application is lacking, maybe because of the lack of quantities required, such as the frictions. In the study of one-dimensional activated barrier-crossing dynamics, we observe that frictional effects can reduce the rate by more than one order of magnitude. Therefore, it is convenient to study the effect of friction on multidimensional barrier-crossing dynamics.[25–27]

The rate of homogeneous nucleation of a liquid droplet in a supersaturated vapor phase at a particular supersaturation value depends on a host of factors. In this study, our system is modeled using Lennard-Jones interactions.[28,29] Even in this simplest of problems, determining the free energy surface as a function of appropriate order parameters (needed to describe the nucleation process) is challenging. This further requires the specification of multiple order parameters.

The present study differs from the scheme of Langer in that we employ a *non-Markovian* multidimensional approach to understand the effects of friction on the nucleation rate.[16,30] Although couched in a different language, the theory is in many respects similar to the multidimensional non-Markovian rate theory developed by Grote and Hynes (MDNMGH). However, MDNMGH does not contain any coupling at the level of Hamiltonian.[31] The two-order parameters are assumed to be orthogonal from the beginning. This coupling is present in our calculations, although one can adopt an orthogonal representation. The orthogonal representation, however, is to be derived from our physically motivated order parameters. The procedure has been described in detail in Section II.

The rest of the paper is organized as follows. Section II recapitulates *the non-Markovian multidimensional theory we adopted in our earlier studies.* In this context, we discuss the theoretical background for the calculation of the barrier crossing rate. In section III, we detail the simulation methodology. Section IV includes the results and discussions. Finally, section V summarizes our work and draws some general conclusions.

## II. MULTIDIMENSIONAL NON-MARKOVIAN RATE THEORY

It was Langer who first advanced the idea that calculation of the rate of nucleation could involve a multidimensional free energy surface and that the rate can be formulated as Kramer's problem of crossing over an activation barrier in the presence of frictional forces.[32] We now follow the approach of Langer to formulate nucleation as a rate process. Subsequently, we employ the Grote-Hynes multidimensional non-Markovian (MDNM) rate theory to calculate the nucleation rate. The first step, of course, is to obtain the nucleation free energy surface as a function of the order parameters described above and quantified below. As discussed above, the two order parameters are given by (i) the total number of liquid-like particles ($n_{liq}$) and (ii) the fluctuation of coordination numbers ($\mu_2$). *The reaction energy path* *is assumed to be given by the calculated minimum energy path from the reactant minimum to the product minimum.* This is a unique path to be determined after the calculation of the free energy surface. However, we find that the order parameters are not orthogonal along the reaction coordinate. We orthogonalize the two order parameters at the reactant well and the barrier by introducing two vectors, **Y** and **Z**, that are orthogonal to each other. We then express the reaction coordinate (**X**) in terms of these two vectors, **Y** and **Z**. Interestingly, our two order parameters are nearly orthogonal near the barrier top. The detailed procedure is provided in section IV. In general, however, the method described above can be implemented whenever the multidimensional free energy surface is available.

In addition to the free energy barrier, we need the frequencies that characterize the reactant well and the barrier region. The cue comes from the one-dimensional description of the reaction energy surface, where we almost always assume the reaction energy surface to be a sum of piecewise harmonic potentials. *The ensuing description of the rate involves three parameters: the free energy barrier, $\Delta G^\dagger$, the reactant well frequency $\omega_X^w$, and the barrier frequency, $\omega_X^b$ along the reaction coordinate **X**.*[33] In such a simple description, the transition state rate is given by

$$k_f^{TST} = \frac{\omega_X^w}{2\pi} \exp\left(-\frac{\Delta G^\dagger}{k_B T}\right) \quad (1)$$

where $k_B T$ is the Boltzmann constant times the temperature. As is well-known, the transition state theory[34] provides an upper bound of the rate constant value as it neglects the frictional effects. The simplest multidimensional description that includes the retarding influence of frictional forces is given by Smoluchowski, which is given as[35]

$$k_f^{SL} = \frac{\omega_X^w \omega_X^b}{2\pi \zeta_{XX}} \exp\left(-\frac{\Delta G^\dagger}{k_B T}\right) \quad (2)$$

where $\omega_X^b$ is the barrier harmonic frequency, and $\zeta_{XX}$ is the friction on the reaction coordinate. The Smoluchowski equation is valid only at large friction, the so-called overdamped limit. A more general expression was provided by Kramers[35], which corrects the mistakes of Smoluchowski at the low friction limit. According to Kramers's theory, the rate is given by

$$k_f^{Kr} = \frac{1}{\omega_X^b}\left[\left(\frac{\zeta_{XX}^2}{4} + (\omega_X^b)^2\right)^{\frac{1}{2}} - \frac{\zeta_{XX}}{2}\right] k_f^{TST} \qquad (3)$$

Here $k_f^{TST}$ is given by Eq.(1). Thus, no extra parameter is needed. However, Kramer's theory fails to predict the reaction rate accurately for the reactions occurring in a highly viscoelastic solvent where frequency-dependent friction plays an important role. Later, Grote-Hynes[36] theory comes out that corrects Kramers theory to include memory effects in the form of frequency-dependent friction.[37]

However, Smoluchowski, Kramers, and Grote-Hynes consider a one-dimensional reaction energy surface and are not appropriate in the present condition. This limitation was removed by Grote and Hynes in a paper published in 1981 and was applied to various problems by van der Zwan and Hynes.[38,39] Although the multidimensional non-Markovian theory was introduced a long time ago, the treatment has not been implemented in many chemical and physical processes. The reason that has already been mentioned is the large number of parameters to include the effects of an orthogonal (to the reaction coordinate) harmonic coordinate on the reaction rate.

We briefly discuss the generalization by Langer of the 1D Kramers theory to an *N*-dimensional surface to study the nucleation rate and is given by the following well-known equation

$$k_f^{KL} = \frac{\lambda_+}{2\pi} \left[ \frac{\det \mathbf{E^w}}{\det \mathbf{E^b}} \right]^{1/2} \exp\left(-\frac{\Delta G^\dagger}{k_B T}\right) \tag{4}$$

Here $\lambda_+$ is the by the positive root of the equation $\det\left(\lambda^2 \mathbf{I} + \lambda \underline{\underline{\zeta}} + \mathbf{E^b}\right) = 0$, which describes the growth rate owing to a slight deviation from the saddle point. $\mathbf{E^b}$ and $\mathbf{E^w}$ are the hessian matrixes at the saddle and the reactant well, respectively. Here $\mathbf{I}$ denotes the unit matrix. $\underline{\underline{\zeta}}$ denotes the frictional tensor that enters through the two-variable Langevin equation. In the overdamped limit, Eq.(4) reduces to the following equation, which is known as Landauer-Swanson equation[40]

$$k_f^{LS} = \left(\frac{\omega_X^b}{\zeta_{XX}}\right) k_f^{TST} \tag{5}$$

Here, the multidimensional transition rate, $k_f^{TST}$ is given by

$$k_f^{TST} = \frac{1}{2\pi} \frac{\prod_{i=1}^{N} \omega_i^w}{\prod_{i=1}^{N-1} \omega_i^b} \exp\left(-\frac{\Delta G^\dagger}{k_B T}\right) \tag{6}$$

Here $\{\omega_i^w\}$ and $\{\omega_i^b\}$ denote the angular frequencies of the stable modes in the reactant well and the saddle, respectively. It is important to note that Langer's treatment for the escape rate is only applicable in the strong coupling limit when nonequilibrium effects due to energy-diffusion-controlled processes can safely be neglected. In deriving Eq.(5), Landauer and Swanson assumed a scalar friction term with zero off-diagonal couplings. Eq.(5) is thus a simple generalization of the Smoluchowski equation.

In the present, generalized, context of nucleation, we adopt a multidimensional non-Markovian theory, discussed and implemented in several of our earlier studies.[16,30] Let us briefly discuss the formalism first. In this model, the motion along the reactive mode is coupled with a number of nonreactive modes. This coupling is accounted for at the Hamiltonian level.

On the other hand, the solvent can dynamically couple the motion along the reactive mode with the nonreactive mode. Therefore, we need to construct a generalized formalism when all modes follow non-Markovian dynamics. At the same time, the reactive mode $X$ is linearly coupled with the $N$-number of nonreactive harmonic modes $\{i\}$, and the Hamiltonian for the system is given by

$$H = \frac{P_X^2}{2M} + U(q_X) + \sum_{i=1}^{N} \left( \frac{p_i^2}{2m_i} + \frac{m_i}{2} \left[ \omega_i^b q_i + \frac{C_i}{m_i \omega_i^b} q_X \right]^2 \right) \tag{7}$$

Here $M$ and $P_X$ denote the mass and the conjugate momentum along the reactive mode $X$. The reaction surface is represented by a double-well potential $U(q_X)$. Near the saddle, we assume $U(q_X)$ to be harmonic and is denoted as, $U(q_X) = U^{\neq} - \frac{1}{2} M \omega_X^{b\,2} q_X^2$. Here $\omega_X^b$ is the harmonic frequency near the saddle along the reactive mode, $U^{\neq}$ is the activation energy, and $q_X$ denotes the location of the barrier along $X$. In Eq.(7), $q_i$ and $p_i$ represent the position and momentum of the $i^{th}$ nonreactive mode near the saddle, respectively. $m_i$ is the mass along the nonreactive mode, $\omega_i^b$ denotes the harmonic frequency associated with the $i^{th}$ nonreactive mode near the saddle, and $C_i$ measures the coupling strength between the reactive mode $X$ and the $i^{th}$ nonreactive mode at the Hamiltonian level. The equation of motion along the reactive mode ($X$) and the nonreactive modes $\{i\}$ can be written as follows [Eqs. (8) and (9)]

$$\ddot{q}_X(t) = (\omega_X^b)^2 q_X(t) - \frac{1}{M} \sum_i \left( C_i q_i(t) + \frac{C_i^2}{m_i (\omega_i^b)^2} q_X(t) \right) - \int_0^t d\tau \zeta_{XX}(\tau) \dot{q}_X(t-\tau) - \sum_i \int_0^t d\tau \zeta_{Xi}(\tau) \dot{q}_i(t-\tau) + R_X(t)$$
(8)

$$\ddot{q}_i(t) = -(\omega_i^b)^2 q_i(t) - \frac{C_i}{m_i} q_X(t) - \int_0^t d\tau \zeta_{ii}(\tau) \dot{q}_i(t-\tau) - \int_0^t d\tau \zeta_{iX}(\tau) \dot{q}_X(t-\tau) + R_i(t) \tag{9}$$

It is important to note that here both $\omega_X^b$ and $\omega_i^b$ are positive. The negative sign before $\omega_i^b$ illustrates the bound oscillatory nature of the nonreactive modes, and the positive sign before $\omega_X^b$ denotes the unstable nature of the reactive mode. $R_X(t)$ and $R_i(t)$ represent the random forces along the reactive and nonreactive modes, respectively. We then perform the adiabatic elimination to rewrite Eq.(8) as

$$\ddot{q}_X(t) = (\omega_X^b)^2 q_X(t) + f(t) - \int_0^t d\tau \zeta_X^{eff}(\tau)\dot{q}_X(t-\tau) + R'_X(t) \tag{10}$$

where the effective friction $\zeta_X^{eff}(t)$ (in the Laplace plane) is defined by,

$$\zeta_X^{eff}(s) = \zeta_{XX}(s) - \sum_i \left[ \frac{C_i \zeta_{Xi}(s)}{M(s^2 + (\omega_i^b)^2 + s\zeta_{ii}(s))} + \frac{s\zeta_{Xi}^2(s)}{s^2 + (\omega_i^b)^2 + s\zeta_{ii}(s)} + \frac{C_i \zeta_{Xi}(s)}{m_i(s^2 + (\omega_i^b)^2 + s\zeta_{ii}(s))} \right]$$

(11)

In Eq.(10), $R'_X(t)$ is the contribution arising from the random noise, and $f(t)$ denotes the contribution coming from the systematic force.[30] However, it is hard to carry out the inverse Laplace transformation of $\zeta_X^{eff}(s)$ to obtain the time-dependent friction. Fortunately, we require only the frequency-dependent friction (in the Laplace plane) for the multidimensional rate calculation, so the above equations suffice. For a multidimensional system following the generalized Langevin equation, the rate is given by

$$k_{GH} = \left(\frac{\lambda_X}{\omega_X^b}\right) k_f^{TST} \tag{12}$$

Here $k_f^{TST}$ is given by Eq.(6), and $\lambda_X$ denotes the frequency with which the reactant molecule passes by Brownian motion through the barrier region. We solve the following self-consistent equation to obtain the reactive frequency ($\lambda_X$)

$$\lambda_X^2 = (\omega_X^b)^2 - \lambda_X \hat{\zeta}_X^{eff}(\lambda_X) + \sum_i \frac{1}{M}\left( \frac{C_i^2}{m_i(\lambda_X^2 + \lambda_X \zeta_{ii}(\lambda_X) + (\omega_i^b)^2)} - \frac{C_i^2}{m_i(\omega_i^b)^2} \right) \tag{13}$$

We simplify Eq.(13) further to obtain

$$\lambda_X = \frac{(\omega_X^b)^2 + \sum_i \frac{1}{M}\left(\frac{C_i^2}{m_i(\lambda_X^2 + \lambda_X \zeta_{ii}(\lambda_X) + (\omega_i^b)^2)} - \frac{C_i^2}{m_i(\omega_i^b)^2}\right)}{\left(\lambda_X + \zeta_X^{eff}(\lambda_X)\right)} \quad (14)$$

It is important to note that under appropriate conditions, our generalized non-Markovian rate can reduce to several well-known theoretical schemes, like Grote-Hynes, Langer, Pollak, etc.[31,32,41]

The frequency-dependent friction given by Eq.(11) without the Hamiltonian coupling, i.e., $C_i = 0$, becomes

$$\zeta_X^{eff}(s) = \zeta_{XX}(s) - \sum_i \left[\frac{s\zeta_{Xi}^2(s)}{s^2 + (\omega_i^b)^2 + s\zeta_{ii}(s)}\right] \quad (15)$$

We get back the effective friction along the reactive mode *X* as derived much earlier by Hynes.[16,31] Under this condition, Eq.(14) becomes

$$\lambda_X = \frac{(\omega_X^b)^2}{\left(\lambda_X + \zeta_X^{eff}(\lambda_X)\right)} \quad (16)$$

We thus recover the self-consistent equation originally derived by Grote-Hynes in the absence of any coupling, which is not surprising. Notably, when there is no coupling at the frictional level, the effective friction becomes independent of the Hamiltonian coupling strength $C_i$. In Eq.(14), we observe that Hamiltonian coupling can alter the reactive frequency, thereby influencing the reaction rate even in the absence of frictional coupling.

However, when frictional coupling is present, Hamiltonian coupling significantly modifies both the effective friction and the reactive frequency, influencing the reaction rate. The preceding discussion on multidimensional rate theory highlights that friction is necessary only in the Laplace plane to obtain the rate. Therefore, in principle, we can calculate it by employing our adopted formalism.

### III.    SYSTEM AND SIMULATION DETAILS

The model system studied consists of particles interacting with each other via Lennard-Jones potential truncated and shifted at different radii ($r_{cut}$)

$$\phi(r) = \phi_{LJ}(r) - \phi_{r_{cut}} \quad ; r \leq r_{cut}$$
$$= 0 \quad\quad\quad\quad ; r > r_{cut}$$

Where $\phi_{LJ}(r) = 4\varepsilon\left[(\sigma/r)^{12} - (\sigma/r)^{6}\right]$ and $r_{cut}$ denotes the cutoff radius for the potential. We carry out biased and unbiased MD simulations in the NPT ensemble using GROMACS software patched with PLUMED.[42] The studied system consists of N=512 argon atoms at temperature T=80.7 K and pressure P=5.0 bar. The potential has been truncated with cutoff at $6.75\sigma$ with $\varepsilon = 0.997 kJ/mol$ and $\sigma = 0.34 nm$. We use the leapfrog algorithm to integrate the equation of motion of the particle with a time step of 2.0 fs. A Berendsen thermostat with a time constant of 0.1 ps is used for temperature coupling. We use Berendsen barostat with a pressure coupling of 2.0 ps to control the pressure. The supersaturation (S) is computed as the ratio of the actual and equilibrium vapor pressure. The equilibrium vapor pressure of Argon at our thermodynamic state is equal to 0.43 bar.[22] Therefore, the supersaturation (S) of the system at this thermodynamic state is found to be 11.6. Later, we tune the pressure of the simulation box to achieve the desired supersaturation level *S*. **Table S1** (Supplementary Material, SM) reports the pressure and corresponding supersaturation used in our study.

In our study, we choose two order parameters, namely (i) the number of liquid-like particles ($n_{liq}$) and (ii) the second moment of the distribution of the coordination numbers ($\mu_2$). In our definition, an atom is identified as liquid if it has more than five neighboring atoms.

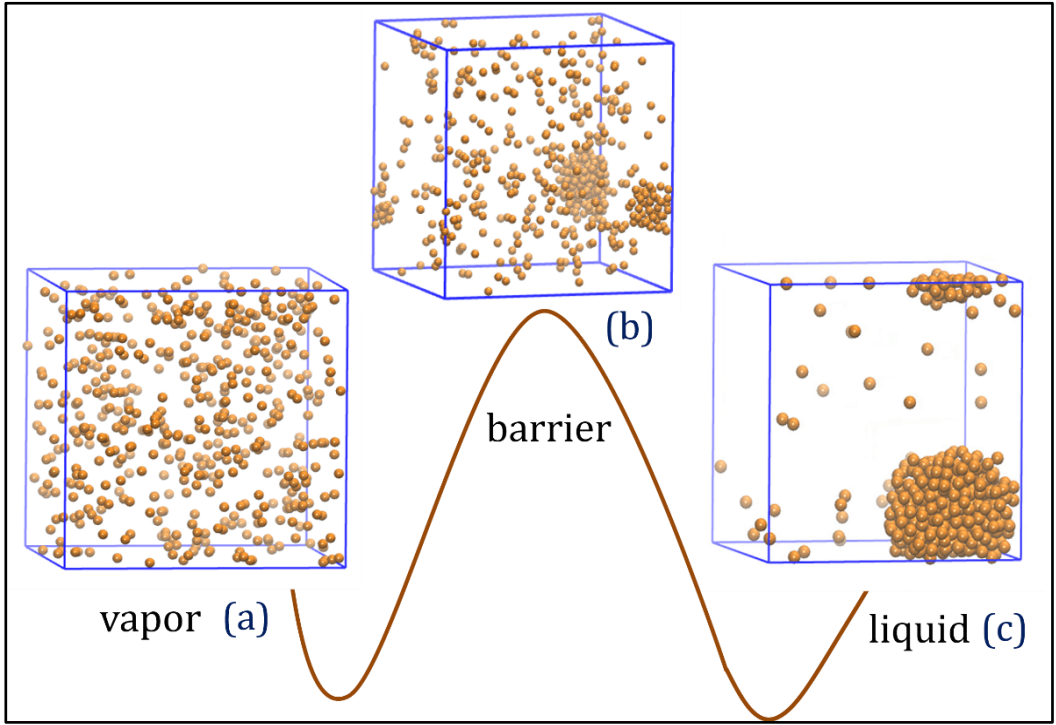

**Figure 1: Snapshots of the system. (a) Initially, particles are randomly distributed inside the cubic box. (b) Snapshot of the system when a cluster just starts to form (c) The snapshot of the liquid phase obtained after a well-tempered metadynamics simulation.**

The coordination number of an atom $i$ is calculated through the use of a switching function as follows [23,43]

$$c_i = \sum_{j \neq i} \frac{1-\left(r_{ij}/r_c\right)^6}{1-\left(r_{ij}/r_c\right)^{12}} \qquad (17)$$

where the summation is performed over all atoms $j \neq i$. The distance $r_{ij}$ between atoms $i$ and $j$ needs to be less than a cutoff $r_c$ to be considered as neighbors. The number of liquid-like atoms $n_{liq}$ can be obtained by using a similar form with threshold value $c_l$, which we take to be 5 in our study

$$n_{liq} = \sum_i \frac{1-\left(c_l/c_i\right)^6}{1-\left(c_l/c_i\right)^{12}} \qquad (18)$$

The above definition of $n_{liq}$ captures the total number of liquid-like atoms in the system. In the context of the problem, it is important to consider the density of the clusters in which these atoms are present if there are more than 1 clusters, the shape of the cluster, etc. To understand these effects, we consider the second moment of the coordination number distribution, which is calculated as $\mu_2 = \frac{1}{N}\sum_{i=1}^{N}(c_i - \bar{c})^2$ where $\bar{c}$ denotes the average of the coordination number.[22] It basically captures the variation in density of the liquid phase and thus can be used to study the local properties of the liquid droplets.

We conduct well-tempered metadynamics simulation to construct the free energy surface as a function of the chosen order parameters for all the studied supersaturation. We can perform unbiased simulation directly in reasonable computer time to study the nucleation for high supersaturation. In this context, we carry out a 200 ns long unbiased MD simulation for high supersaturation since it can capture the nucleation event effectively. In **Figure S1** (SM), we show the variation of these order parameters with time obtained from unbiased simulation.

## IV. RESULTS & DISCUSSIONS

### a. Calculation of free energy surface

We have already discussed why we selected the two order parameters (the liquidness of the system and the fluctuation in the coordination number of liquid-like clusters) to understand the gas-liquid nucleation phenomenon. All the results we present now are for fixed supersaturation, i.e., S=11.6. In **Figure 2a**, we present the free energy surface calculated using well-tempered metadynamics simulation as a function of two order parameters, $n_{liq}$ and $\mu_2$.[22] After we obtain the free energy surface, the next step is to obtain the reaction path. The latter is defined by the minimum energy pathway on the two-dimensional free energy surface, the $n_{liq}$-$\mu_2$ plane. **Figure 2b** shows the most probable minimum energy pathway in the two-dimensional $n_{liq}$-$\mu_2$ plane by employing the MEPSA algorithm.[44] In **Figure 1**, we show the snapshots of the most probable configurations near the reactant and product well, as well as the barrier regime during the gas-liquid nucleation. In **Figure 6a**, we show some unbiased trajectories on the free energy surface, validating the minimum energy pathway drawn in **Figure 2b**. The chosen order parameters (i.e., $n_{liq}$ and $\mu_2$) effectively capture the nucleation process, as found by the analysis of the evolution of the cluster sizes, in particular, the liquid-like clusters. However, they are not fully orthogonal. This is an interesting and could be a common condition along the reaction path. When the two coordinates are not orthogonal, we face complexity in the calculations of the rate, as the equations of motions become coupled and off-diagonal frictions become hard to define. Therefore, we orthogonalize the two order parameters by introducing two mutually orthogonal vectors, **Y** and **Z**, as a linear combination of $n_{liq}$ and $\mu_2$. Following the Schmidt orthogonalization procedure, we set $\mathbf{Z} = an_{liq}\hat{\mathbf{j}}_{\mathbf{n}_{liq}} + b\mu_2\hat{\mathbf{j}}_{\mu_2}$ and $\mathbf{Y} = bn_{liq}\hat{\mathbf{j}}_{\mathbf{n}_{liq}} - a\mu_2^2\hat{\mathbf{j}}_{\mu_2}$ where $a$ and $b$ are the coefficients. Here $\hat{\mathbf{j}}_{\mathbf{n}_{liq}}$ and $\hat{\mathbf{j}}_{\mu_2}$ are the unit vectors along $n_{liq}$ and $\mu_2$ coordinates, respectively. The reaction coordinate $X$ can now be expressed in terms of these two mutual orthogonal vectors in the $Y$-$Z$ plane. However, our present study finds that the reaction coordinate (denoted by $X$) completely aligns with the **Z**-direction near the barrier top. Therefore, **Y**, which is orthogonal to the reaction coordinate (**X**), serves as a nonreactive mode in our case. In this work, we only calculate the rate along the collective coordinates **X** and **Y**, but they retain all the information.

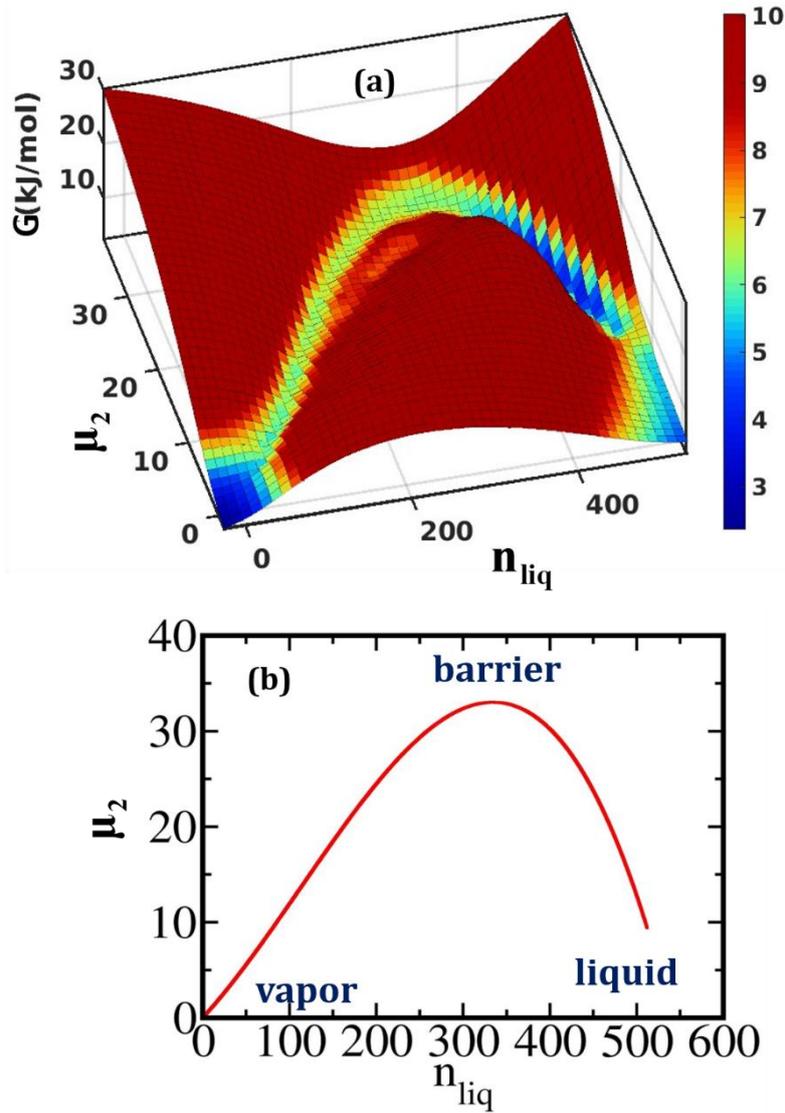

**Figure 2: (a) The computed three-dimensional free energy surface of gas-liquid nucleation as a function of the two collective variables $n_{liq}$ (number of liquid-like atoms) and $\mu_2$ (the second moment of the coordination number distribution) at S=11.6. In (b), the red line indicates the minimum energy pathway connecting the initial vapor and final liquid states during the nucleation at S=11.6.**

**b.   Well and barrier frequencies and frequency-dependent friction**

We now discuss the theoretical formalism to determine the reactant well and the barrier frequencies, the frequency-dependent friction, and the effective mass. The formalism prescribed by Hynes[45,46] is followed to calculate reactant well frequencies, the friction, and the effective mass of the system bounded in a harmonic well. Let us consider a harmonically bound tagged particle is connected with a heat bath at temperature *T*. In this case, the tagged particle experiences a restoring force $-\mu\omega^2 X$ (here $\mu$ denotes the effective mass, $\omega$ is the harmonic frequency, and *X* is the reaction coordinate). In this scenario, one can start with the Langevin equation to derive the following equations[47] (see Chandrasekhar) for the equilibrium position and the velocity [Eq.(19) and Eq.(20)]

$$\langle \delta \mathbf{X}(0)^2 \rangle = \frac{k_B T}{\mu \omega^2} \tag{19}$$

and

$$\langle \delta \dot{\mathbf{X}}(0)^2 \rangle = \frac{k_B T}{\mu} \tag{20}$$

Here $\mu$ is the effective mass of the collective variable *X* and $\omega$ is the well frequency along the collective variable *X*. By running well-specific unbiased simulations, we can directly calculate $\langle \delta \mathbf{X}(0)^2 \rangle$ and $\langle \delta \dot{\mathbf{X}}(0)^2 \rangle$. Next, we estimate $\mu$ and $\omega$ in the following way

$$\omega = \sqrt{\frac{\langle \delta \dot{\mathbf{X}}(0)^2 \rangle}{\langle \delta \mathbf{X}(0)^2 \rangle}} \tag{21}$$

And

$$\mu = \frac{k_B T}{\langle \delta \dot{\mathbf{X}}(0)^2 \rangle} \tag{22}$$

After having the two well frequencies, our next goal is the calculation of frequency-dependent friction. In this regard, we define a normalized time correlation function as $C_{XX}(t) = \frac{\langle \mathbf{X}(0)\mathbf{X}(t) \rangle}{\langle \mathbf{X}(0)^2 \rangle}$. To calculate friction on *X*, we write down the generalized Langevin equation for $C_{XX}(t)$ and simplify it in the frequency (s) plane to obtain[47]

$$C_{XX}(s) = \frac{s + \zeta_{XX}(s)}{s^2 + \omega^2 + s\zeta_{XX}(s)} \tag{23}$$

where $C_{XX}(s) = \int_0^\infty dt\, e^{-st} C_{XX}(t)$ and $\zeta_{XX}(s)$ is the frequency-dependent friction along the reactive coordinate *X*. It is convenient to approximate an exponential decay for $\zeta_{XX}(t)$, i.e., $\zeta_{XX}(t) = \zeta_{XX}(0)\exp(-t/\tau)$. Here $\tau$ is the decay constant associated with the exponential relaxation and $\zeta_{XX}(0)$ denotes the friction at *t*=0. The Laplace transformation provides the frequency-dependent friction $\zeta_{XX}(s)$, i.e., $\zeta_{XX}(s) = \int_0^\infty dt\, e^{-st}\zeta_{XX}(t) = \frac{\zeta_{XX}(0)}{s + 1/\tau}$. We substitute it in Eq.(23), and obtain an expression for $C_{XX}(s)$ as a function of *s*. We then numerically fit the plot of $C_{XX}(s)$ against *s* obtained via simulation to estimate $\zeta_{XX}(0)$ and $\tau$.

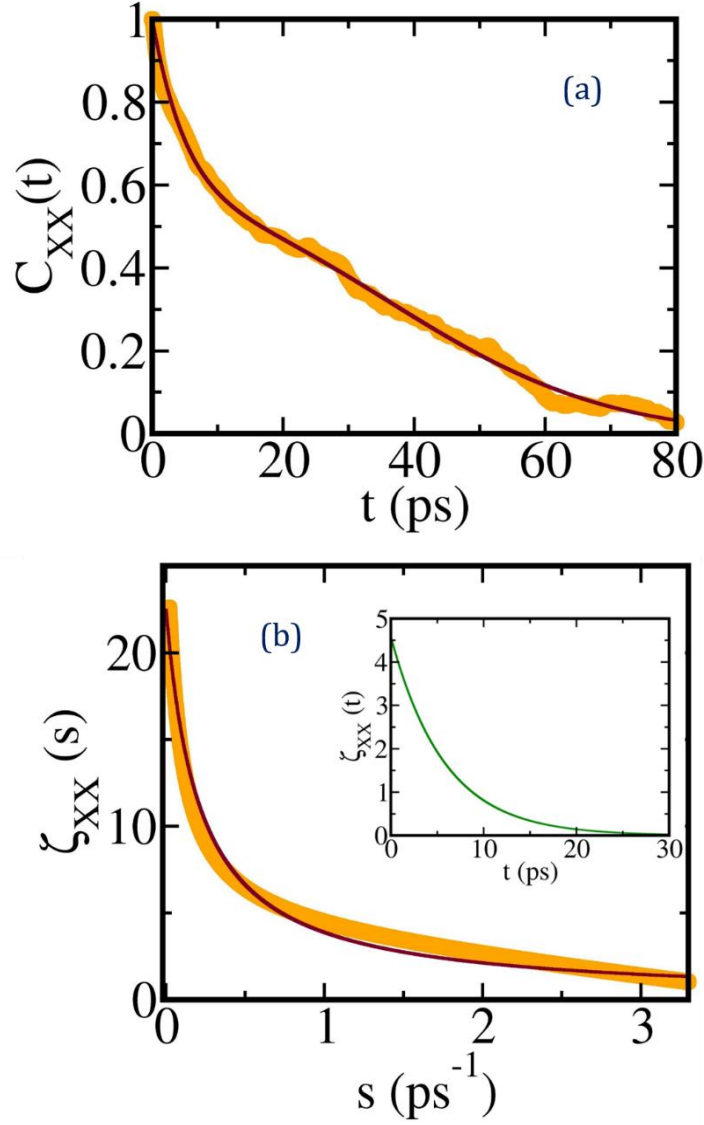

**Figure 3.** (a) The plot of the normalized time-correlation function $C_{XX}(t)$ against t, as shown by the yellow line at S=11.6. In (a), we use the function $a_0 \exp(-t/\tau_1) + a_1 \exp(-t/\tau_2) + a_2 \exp(-t/\tau_3)$ to fit the time correlation function shown by the dark brown line since the selected fitting function produces the maximum fitting correlation coefficient (i.e., 0.999) over other functions. (b) Frequency dependence of the friction on reaction coordinate against frequency s at S=11.6. We carry out a Laplace transformation to obtain $C_{XX}(s)$ and invoke Eq.(23) to generate frequency-dependent friction $\zeta_{XX}(s)$. We employ the function $\frac{\zeta_{XX}(0)}{(s+1/\tau)}$ to fit the frequency-dependent friction, as shown by the dark brown line. The inset in (b) displays the time-dependent friction obtained after the inverse Laplace transformation.

As discussed in the previous section, to avoid the off-diagonal terms, we perform a coordinate transformation from the $n_{liq}$-$\mu_2$ plane to the mutually orthogonal $X$-$Y$ plane using the Gram-Schmidt orthogonalization technique, where $X$ denotes the reaction coordinate.[48] We compute the fluctuation in $\mathbf{X}(t)$ and construct a time-correlation function $\langle \mathbf{X}(0).\mathbf{X}(t) \rangle$ that contains four terms. If $\mathbf{X} = a n_{liq} \hat{\mathbf{j}}_{n_{liq}} + b \mu_2 \hat{\mathbf{j}}_{\mu_2}$ then,

$$\langle \mathbf{X}(0).\mathbf{X}(t) \rangle = a^2 \langle n_{liq}(0) n_{liq}(t) \rangle + b^2 \langle \mu_2(0) \mu_2(t) \rangle + ab \left[ \langle n_{liq}(0) \mu_2(t) \rangle + \langle \mu_2(0) n_{liq}(t) \rangle \right]$$

We perform multiple unbiased MD simulations in the initial vapor state with a 1fs data dumping resolution and monitor the two order parameters, $n_{liq}$ and $\mu_2$, as discussed earlier. The distributions of $n_{liq}$ and $\mu_2$ show that the system remains in the reactant well, and then we proceed to the calculation of the time correlation function. In **Figure 3(a)**, we plot the normalized time-correlation function $\langle \mathbf{X}(0).\mathbf{X}(t) \rangle$ of the collective variable $X$ against $t$. We invoke Newton's difference quotient method to calculate the numerical time derivative of $X$ with a minimal value of $\delta t$ (i.e., 0.001) to make the calculation robust. We estimate $\langle \delta \mathbf{X}(0)^2 \rangle$ and $\langle \delta \dot{\mathbf{X}}(0)^2 \rangle$ to obtain the well frequency using Eq.(21). In **Figure 3(b)**, we plot frequency-dependent friction against frequency (*s*) and evaluate the frequency-dependent friction along $X$ by employing Eq.(23).

The entire procedure outlined above for the $X$-coordinate is repeated for the $Y$-coordinate.

We identify several barrier-top conformations from the metadynamics trajectory and subsequently conduct short, unbiased simulations, starting at the barrier-top. Following this, we stitch these short trajectories and compute the mean squared fluctuations near the barrier. We repeat this process to determine both the barrier frequencies and frequency-dependent friction along both reactive and nonreactive coordinates.

### c. Rate of Nucleation

We have mentioned earlier that an accurate calculation of nucleation rate is a formidable problem, and to the best of our knowledge, only a few such calculations exist. In the following, we demonstrate that the present theoretical method provides a fairly good estimate of nucleation, as evident from comparison with the rates obtained from direct unbiased MD simulations that is possible at large supersaturations S. We note here that many of the earlier simulations on the spinodal point of nucleation underestimate the value of limiting S because they used a truncated form of potential. In most of the simulations of gas-liquid nucleation with Lennard Jones, the potential was truncated at the cutoff distance of 2.5σ.[8,21] In the present simulations, spinodal $S$ is about 15, as discussed in the subsequent section. We should stress that the value of the spinodal depends on the system size and thermodynamic state.

In **Table 1**, we calculate the required parameters for calculating the rate using the formalism described above at S=11.6. Next, we calculate the rate constants according to the different theoretical prescriptions by using these parameters as input. We invoke Kramers-Langer's theory[24,49] and the simpler theory of Landauer and Swanson[40] to calculate the nucleation rate in order to understand the effect of the multidimensional nature of the free energy surface on the reaction rate. However, these formalisms do not consider the memory effects. This is why we invoke the multidimensional non-Markovian theory adopted in our study to calculate the nucleation rate.

**Table-1: The required parameters for the calculation of nucleation rate along the mutually orthogonal coordinates, *X* and *Y*. We have calculated the parameters by employing Hynes formalism at S=11.6.**

| Well frequency along X ($\omega_X^w$) | $\omega_X^w = (1.31 \pm 0.02) ps^{-1}$ |
|---|---|

| **Well frequency along Y ($\omega_Y^w$)** | $\omega_Y^w = (1.30 \pm 0.01)\, ps^{-1}$ |
|---|---|
| **Barrier frequency along X ($\omega_X^b$)** | $\omega_X^b = (1.84 \pm 0.02)\, ps^{-1}$ |
| **Barrier frequency along Y ($\omega_Y^b$)** | $\omega_Y^b = (1.56 \pm 0.01)\, ps^{-1}$ |
| **Zero frequency Friction along X, $\zeta_{XX}(0)$** | $\zeta_{XX}(0) = (26.44 \pm 0.01)\, ps^{-1}$ |
| **Time-dependent friction, along X** | $\zeta_{XX}(t) = 4.58 \exp\left(-\dfrac{t}{(5.77 \pm 1.0)/ps}\right)$ |
| **Friction along Y, $\zeta_{YY}(0)$** | $\zeta_{YY}(0) = (26.24 \pm 0.01)\, ps^{-1}$ |
| **Free energy barrier ($\Delta G^\dagger$)** | $\Delta G^\dagger = 4.56\, kJ/mol$ |

In **Table 2**, we provide the numerical values for rate constants along the reactive mode, *X*. We employ Eq.(14) to calculate the reactive frequency $\lambda_X$ for calculating the rate using non-Markovian theory. We estimate $\zeta_{XX}$ and $\zeta_{YY}$ from unbiased simulation employing Hynes formalism. Estimating coupling friction $\zeta_{XY}$ between the reactive and nonreactive coordinates poses a challenge. The conventional method for calculating off-diagonal friction is particularly challenging in this context, as the technique employed for diagonal elements proves ineffective. As noted by vdZ-H, off-diagonal coupling friction arises due to the influence of solvent fluctuations. By utilizing the mode-coupling theory, we can establish a connection with the dynamic structure factor of the surrounding water molecules. Implementing such a theoretical approach involves the use of Density Functional Theory. However, it is extremely tricky to perform a detailed analysis of this at this stage.

**Table-2:** The numerical values of rate constants along the reaction coordinate $X$ at $S=11.6$. Here, we have calculated the rate from different theoretical approaches. The appropriate parameters used here are given in Table 1.

| Theoretical approach | Expressions | Numerical value |
|---|---|---|
| **1D transition state theory** | $k_f^{TST} = \dfrac{\omega_X^w}{2\pi} \exp\left(-\dfrac{\Delta G^\dagger}{k_B T}\right)$ | $(232.66 \pm 1.0)\,\mu s^{-1}$ |
| **1D Kramers' theory** | $k_f^{Kr} = \dfrac{1}{\omega_X^b}\left[\left(\dfrac{\zeta_{XX}^2}{4} + (\omega_X^b)^2\right)^{\frac{1}{2}} - \dfrac{\zeta_{XX}}{2}\right] k_f^{TST}$ | $(16.11 \pm 0.1)\,\mu s^{-1}$ |
| **1D Smoluchowski equation** | $k_f^{SL} = \left(\dfrac{\omega_X^b}{\zeta_{XX}}\right) k_f^{TST}$ | $(16.19 \pm 0.1)\,\mu s^{-1}$ |
| **2D Transition state theory** | $k_{2D}^{TST} = \left(\dfrac{\omega_Y^w}{\omega_Y^b}\right) k_f^{TST}$ | $(193.88 \pm 0.11)\,\mu s^{-1}$ |
| **Kramers-Langer** | $k_f^{KL} = \dfrac{\lambda_+}{2\pi}\left[\dfrac{\det \mathbf{E}^w}{|\det \mathbf{E}^b|}\right]^{1/2} \exp\left(-\dfrac{\Delta G^\dagger}{k_B T}\right)$ | $(13.38 \pm 0.2)\,\mu s^{-1}$ |
| **Grote-Hynes theory with non-Markovian friction along both modes (with zero coupling friction)** | $k_{GH} = \left(\dfrac{\lambda_X}{\omega_X^b}\right) k_{2d}^{TST}$ | $(32.10 \pm 0.2)\,\mu s^{-1}$ |
| **Landauer and Swanson** | $k_f^{LS} = \left(\dfrac{\omega_X^b}{\zeta_{XX}}\right) k_{2d}^{TST}$ | $(13.39 \pm 0.1)\,\mu s^{-1}$ |

**d.    Interplay between dimensionality and memory effects**

Results presented in **Table 2** reveal two notable trends. Initially, as we transition from a one-dimensional framework to a two-dimensional one, a marked decrease in the reaction rate is evident across various instances such as TST, GH, etc. Simultaneously, it becomes apparent that, in the one-dimensional context, the rate associated with Markovian-based Kramers' theory is considerably lower than the Grote−Hynes rate constant. The interplay between dimensionality and memory effects elucidates these conflicting trends.[50]

Grote−Hynes extended Kramers' theory by eliminating the white noise assumption to encompass non-Markovian effects. The significance of non-Markovian/memory effects is underscored in reactions featuring sharp, high-frequency barriers within a highly viscoelastic solvent medium. Here, the slow motions coming from the viscous solvent fail to couple with barrier-crossing dynamics. Consequently, friction at the zero-frequency limit overestimates the actual friction, necessitating frequency-dependent friction for accurate rate estimation. Consequently, Grote-Hynes theory predicts a higher rate than Kramers' rate in one dimension.

Another pivotal factor significantly influencing the reaction rate is the dimensionality of the reaction energy surface. Notably, the involvement of an additional dimension consistently diminishes the reaction rate akin to viscosity effects. This observation is attributed to the increased time available for the dissipation of excess kinetic energy in the multidimensional scenario. Consequently, we can deduce that an extra dimension curtails the reaction rate by increasing friction along the reaction coordinate. Therefore, it is evident that both opposing factors—dimensionality and memory effects—are equally crucial in our comprehensive analysis.

**e.    Supersaturation dependence of free energy barrier and nucleation rate**

In the preceding section, we calculated the nucleation rate using both multidimensional Markovian and non-Markovian theories at a supersaturation level of 11.6.

This section explores the sensitivity of the theoretical nucleation rate to changes in supersaturation. To investigate this, we vary the supersaturation by adjusting the pressure of the system. In addition to S=11.6, we consider three more supersaturation levels, namely S=7.0, S=10.4, and S=12.8, in our calculations.

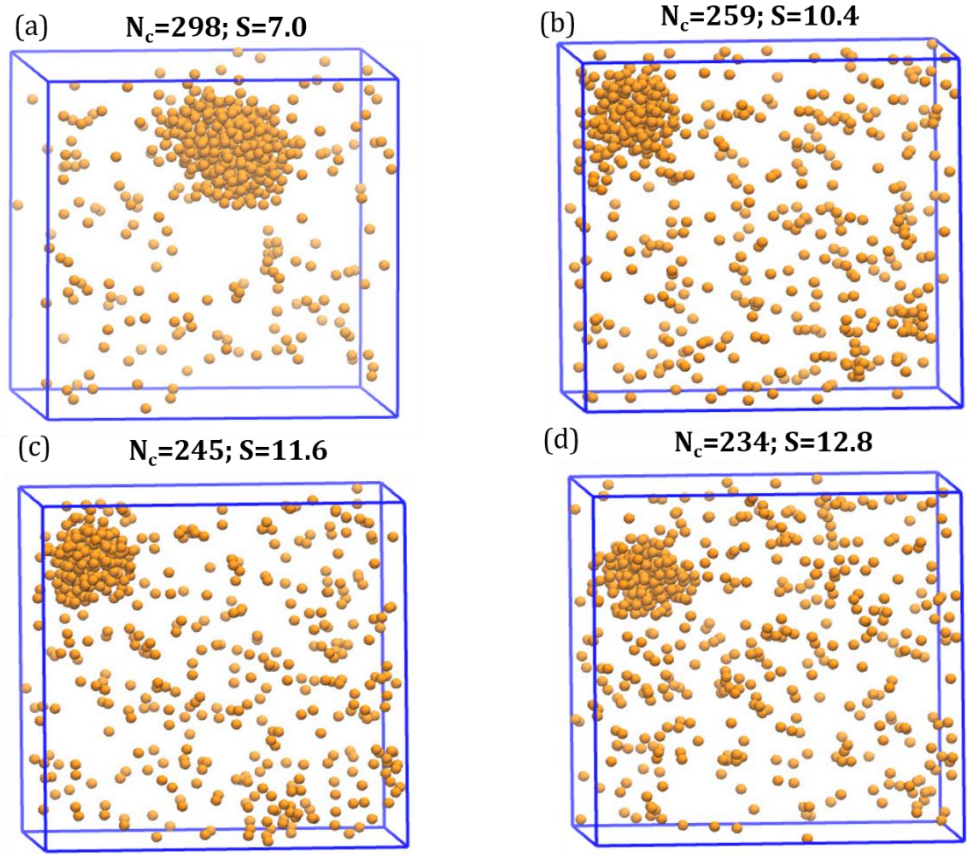

**Figure 4: Snapshots of the largest clusters near the barrier top at different supersaturations. With increasing supersaturation, the size of the critical clusters decreases.**

In **Figure 4**, we present the snapshots of the largest cluster near the barrier top at various supersaturations. The figure shows that the size of the largest cluster ($N_c$) near the barrier top decreases with increasing supersaturation, resulting in a significant rise in the nucleation rate as supersaturation increases. Following a similar protocol, we initially construct the free energy surface as a function of the two order parameters (i.e., $n_{liq}$ and $\mu_2$) for the system with different supersaturations. Next, we orthogonalize the coordinates to avoid the off-diagonal coupling.

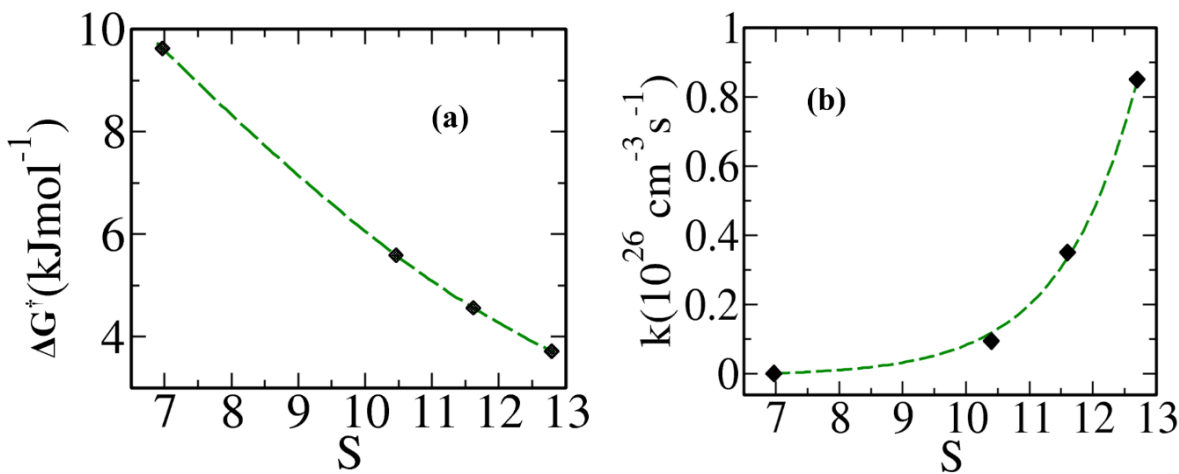

**Figure 5: (a) Free energy barrier against the supersaturation. With increasing supersaturation, the free energy barrier decreases, suggesting that nucleation becomes more favorable with increasing supersaturation. (b) Rate of the nucleation against supersaturation. Rates shown in the plot are obtained from the multidimensional non-Markovian rate theory adopted here. In (a), the green dashed lines joining the data points are provided as a guide to the eyes. In (b), we perform an exponential fit for the data points, as shown by the green dashed line.**

We report the necessary parameters for three distinct supersaturation values, including well frequencies, barrier frequencies, frequency-dependent friction, etc., in SM (see SM-S3, SM-S4, and SM-S5 for details). Our findings indicate that the barrier frequency gradually decreases with increasing supersaturation, suggesting a flattening of the barrier. In **Figure 5a**, we illustrate the variation of the free energy barrier against supersaturation, observing a decrease in the nucleation barrier with higher supersaturation. This implies that nucleation becomes more favorable as the supersaturation increases.

The nucleation rates predicted by the Markovian and non-Markovian formalisms are provided in the Supplementary Material (i.e., see SM-S3, SM-S4, and SM-S5) for S=7.0, S=10.4 and S=12.8. **Figure 5b** depicts the nucleation rate predicted by the multidimensional non-Markovian rate theory, revealing a significant dependence on supersaturation. This figure shows a divergence-like growth in the nucleation rate, as observed in experiments.[10]

To validate our theoretical findings, it is convenient to obtain the nucleation rate directly from simulations. Accordingly, we employ the Mean First Passage Time (MFPT) formalism, as discussed in the following section. Initially, estimating the nucleation rate using the MFPT formalism at S=7.0 faced challenges via an unbiased simulation due to the high free energy barrier. Nevertheless, in the following section, we explore the rate calculation via the MFPT formalism for three other supersaturations, namely S=10.4, 11.6, and 12.8, where nucleation can be observed even in an unbiased simulation.

**f.  Calculation of nucleation rate from MD simulation via Mean first-passage time method (MFPT)**

There are several ways one can calculate nucleation rate directly from simulation.[11,22] We employ mean first passage time (MFPT) formalism to calculate the nucleation rate in this work. In the case of nucleation, MFPT refers to the average time $\tau(n)$ that a cluster takes to reach a particular size *n* for the first time. We calculate the time until the largest cluster in the system reaches or exceeds any given cluster size *n* for the first time and take an average over several trajectories starting from the vapor phase to obtain the mean first passage time. The plot of MFPT against the size of the cluster (*n*) usually exhibits a characteristic sigmoidal shape, and in the case of a sufficiently high nucleation barrier, it can be demonstrated accurately by the following expression $\tau(n) = \frac{\tau_{MFPT}}{2}\left[1 + erf\left(b(n - n^*)\right)\right]$ where $erf(x) = \frac{2}{\sqrt{\pi}}\int_0^x dt e^{-t^2}$ is the error function and *b* is related to the Zeldovich factor, i.e., $Z = b/\sqrt{\pi}$.[11] We can now calculate the nucleation rate from nucleation time $\tau_{MFPT}$ as $J = \frac{1}{V\tau_{MFPT}}$. Here *V* is the volume of the system.

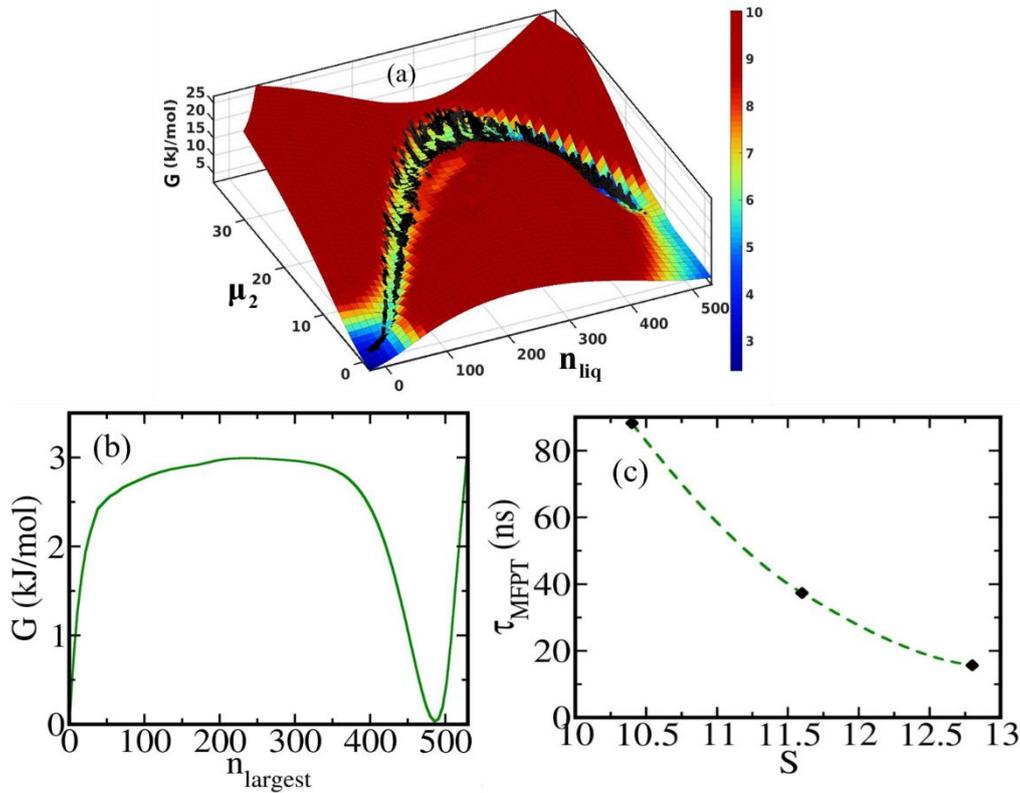

**Figure 6: (a) Nucleation trajectories (as shown by black colored lines) generated from unbiased simulations on the three-dimensional free energy surface starting from the vapor phase at S=11.6. The figure shows the trajectories to follow the minimum energy pathway. (b) The plot of the free energy against the size of the largest cluster size at S=11.6 obtained from unbiased simulation. (c) MFPT as a function of supersaturation. Here, the green dashed lines joining the data points are provided as a guide to the eyes Clearly, nucleation time decreases with increasing supersaturation.**

In the nucleation process, studying the largest liquid cluster is important. We use the geometric cluster criteria that are derived from the one proposed by Stillinger to identify the particles that constitute the largest cluster. [23,43] All the liquid-like particles that are less than $q_c = 1.5\sigma$ apart are considered to be connected and, therefore, belong to the same cluster. We conduct numerous unbiased simulations, initiating from the vapor phase, and record the free energy profile against the size of the largest cluster (depicted in **Figure 6b**). Two trajectories are shown on the free energy surface obtained from the metadynamics simulation by the black lines in **Figure 6a** at S=11.6. Using the abovementioned procedure, we calculate the Mean First Passage Time (MFPT) for different supersaturations and plot it against the supersaturation in **Figure 6c**. From the figure, we find that the system takes less time to reach the cluster of threshold size as we increase the supersaturation. A comparison of the nucleation rates obtained from both theoretical predictions and simulations is presented in **Table 3**.

**Table-3: Comparison of nucleation rates obtained from Markovian and non-Markovian theories and simulation for different supersaturations. For S=7.0, it is not possible to estimate the nucleation rate directly from unbiased simulation because of the high free energy barrier.**

| Supersaturation (S) | From Markovian Theory (Langer) ($cm^{-3}s^{-1}$) | From Non-Markovian theory ($cm^{-3}s^{-1}$) | From Simulation via MFPT ($cm^{-3}s^{-1}$) |
|---|---|---|---|
| 7.0 | $(0.11 \pm 0.03) \times 10^{23}$ | $(0.20 \pm 0.03) \times 10^{23}$ | ------------------------ |
| 10.4 | $(0.04 \pm 0.01) \times 10^{26}$ | $(0.09 \pm 0.02) \times 10^{26}$ | $(0.11 \pm 0.02) \times 10^{26}$ |
| 11.6 | $(0.13 \pm 0.02) \times 10^{26}$ | $(0.32 \pm 0.05) \times 10^{26}$ | $(0.26 \pm 0.07) \times 10^{26}$ |
| 12.8 | $(0.42 \pm 0.05) \times 10^{26}$ | $(0.81 \pm 0.10) \times 10^{26}$ | $(0.62 \pm 0.10) \times 10^{26}$ |

We observe a strong agreement between the simulation results and theoretical predictions for different values of supersaturation. The rate obtained by the non-Markovian theory is marginally higher than that predicted by the Markovian theory, highlighting the influence of the memory effect in the process. It is important to note that our findings align with other theoretical results, particularly those obtained from unbiased simulations utilizing different methods such as direct flux, survival probability, MFPT, etc.[11,22]

g.    **Estimation of the kinetic spinodal decomposition point**

In the well-known van der Waals picture, the spinodal decomposition point in a gas-liquid transition is determined by the extrema of the van der Waals loop and is given by the divergence of the isothermal compressibility.[1,2] This is known as the thermodynamic definition of the spinodal decomposition point, where the homogeneous gas (or liquid) becomes unstable to infinitesimal density fluctuations. As demonstrated in earlier studies,[7,20] this is manifested in the simulations as the appearance of large-scale density fluctuations. There is also a kinetic definition of a spinodal point, which is the point where the nucleation rate diverges.

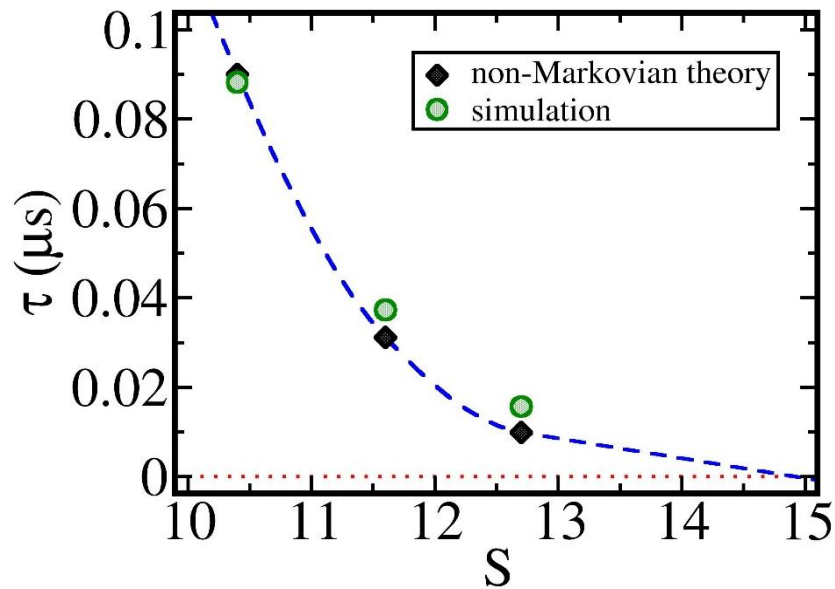

**Figure 7: Plot of nucleation time against supersaturation. The nucleation time, represented by black diamonds, is determined by taking the inverse of the rate obtained from the non-Markovian theory. For reference, the mean first-passage time (MFPT) is depicted with green circles. To identify the kinetic spinodal point, we perform a cubic spline fitting for extrapolation, displayed by the blue dashed line. The red dotted line corresponds to the zero nucleation time. In this particular system, the spinodal is found to occur at S=14.8, where the nucleation rate diverges.**

We find the signature of this kinetic spinodal in our calculations. Specifically, we determine through extrapolation that the kinetic spinodal occurs at S=14.8 in the current system. The dependence of the nucleation time (inverse of the rate) against S is plotted in **Figure 7**. Notably, the behavior of the nucleation time at smaller supersaturation S differs from that observed when approaching the spinodal closely.

As discussed earlier, the nucleation rate depends on the thermodynamic conditions. Typically, under the usual experimental conditions characterized by not too high supersaturation or a high degree of supercooling, the rates are slower than the ones observed in simulations. Under such conditions, the height of the nucleation barrier falls within the range of 30-60 $k_BT$, giving a rate in the range of $10^5$-$10^9$ cm$^{-3}$s$^{-1}$.[10] This is the range of time scales usually accessible by the experimental techniques employed to measure the nucleation time.[1] Typically, the cluster has to grow beyond a certain size to be detected by fluorescence or other detection methods. It's crucial to note that the barrier height is not only dependent on supersaturation S but also sensitive to the size of the system under investigation.[13]

However, computer simulations face several constraints, including a finite system size (volume ~$10^{-17}$ cm$^{-3}$) and short simulation time (~$10^{-8}$ s). If a nucleation event is to be observed within these timeframes, it needs to be at very large supersaturation, almost close to the spinodal, where the "measured" rate of nucleation is expected to be around $10^{25}$ cm$^{-3}$s$^{-1}$. Indeed, the prediction made by the multidimensional non-Markovian rate theory for nucleation on the free energy surface calculated by biased simulations does predict small rates observed in experiments.

*Furthermore, in most earlier studies, a truncation of the Lennard-Jones (LJ) potential was employed with a cut value of 2.5 $\sigma$.*[8,21,43,51] *Subsequent study has demonstrated that such truncation leads to a significant discrepancy near the phase transition point.*[51,52] Therefore, in the present study, we purposely avoided employing a cutoff and obtained the nucleation rate.

Notably, these values are in good agreement with earlier studies conducted by Reguera and coworkers and those obtained from our simulations.[11] The historical examination of the rate of gas-liquid nucleation has remained highly controversial. The current study, utilizing two-order parameters, appears to provide a consistent description and also offers an explanation for the mentioned discrepancies.

The simulation of gas-liquid nucleation becomes particularly complex due to the presence of finite-size effects.[13] Finite size effects play a crucial role in determining gas-liquid nucleation, as explained below. This barrier is influenced by various factors, including supersaturation, solvent, temperature, and the entire cluster size distribution.[53] Importantly, the cluster size distribution is influenced by temperature, volume, or the system size. It is only as the system size approaches a macroscopic limit that the cluster size distribution transforms into an intensive property. Our preliminary simulations have indeed shown a significant dependence of the free energy barrier on small system sizes (see **Figure S2** in the Supplementary Material).

## V. CONCLUSIONS

Nucleation is a fundamental physical process that has drawn substantial interest over the years. Although, the classical nucleation theory (CNT) of Becker-Doring-Zeldovich is popularly discussed in textbooks because of its simplicity, it contains questionable assumptions and is found to be grossly inadequate.[4,6] The general theory of Langer was expected to provide a better description but has hardly been used.

In this study, we implement a multidimensional non-Markovian (MDNM) rate theory presented elsewhere to obtain the rate after obtaining the free energy surface by biased molecular dynamics simulations. The rate theory is essentially an adaptation of the multidimensional Grote-Hynes theory, although inspired by Langer and also partly by Pollak. We believe that this combination of multidimensional free energy calculation with MDNM rate theory is a positive step forward.

Here, we report the calculation of gas-liquid nucleation primarily at relatively large supersaturation, S (S is defined as the ratio of pressure $P$ of the system to its coexistence pressure). We find the generalized theory proposed here is quantitatively accurate. In fact, the agreement with the simulated values at multiple values of S is perhaps among the best theoretical estimates that have been achieved to date. We have confined our studies to the values of supersaturation S equal to 10 and above because the nucleation rate is low at small S values, and the process is not observed in unbiased MD simulations. We also extend our analysis at low S, where nucleation can't be observed in unbiased simulation. We also identify the kinetic spinodal, where the gas transforms to liquid immediately and spontaneously, as there is no nucleation barrier.

Let us summarize the steps involved and the results obtained. First, we select two order parameters that characterize the growth of clusters. Here, we select $n_{liq}$ and $\mu_2$ as the two parameters that characterize the system. We next carry out biased MD simulations to obtain the two-dimensional free energy surface. The calculated free energy provides the value of the critical cluster. In the third step, we obtain the minimum energy pathway in this two-dimensional surface, which serves as our reaction coordinate in the calculation of the rate. In the fourth step, we orthogonalize the order parameters by using two new coordinates, **Z** and **Y**, which are orthogonal to each other. We express the primary reaction coordinate $X$ in terms of these two orthogonal vectors. In the 5$^{th}$ step, we obtain the relevant harmonic frequencies, both at the reactant well and the barrier. Subsequently (6$^{th}$ step), we obtain the frequency-dependent frictions. And finally, we obtain the rate from the multidimensional non-Markovian rate theory (MDNMRT). We also obtain the rate from direct unbiased MD simulation that is accessible at large values of S (but not at small values of S). Our study demonstrates that the adopted non-Markovian theory performs reasonably well across various supersaturation values. Furthermore, it exhibits good agreement with earlier theoretical findings.[11,22]

The theoretical scheme developed here is fairly involved. It is to be noted that after the calculation of the free energy surface, our method still requires short, unbiased MD simulations both at the reactant well and at the barrier or saddle surface to obtain the necessary parameters. The scheme is particularly useful at low values of S where unbiased MD simulations cannot give the rate. We find that the transition state theory estimate breaks down -- it overestimates the rate by more than one order of magnitude.

It is interesting to place the effects of friction on nucleation rate in the context of the Zeldovich treatment, where both addition to and evaporation of molecules from the critical nucleus are allowed. This evaporation allows the back-recrossing of the barrier and leads to the breakdown of the transition state theory, as observed here. Just as in Zeldovich treatment, barrier frequency also plays an important role here. The main difference is that we have adopted non-Markovian kinetics on an accurately determined multidimensional free energy surface. The primary difficulty with Zeldovich treatment is the use of capillary approximation. It was shown later that both the critical cluster size and the free energy barriers are substantially underestimated. For a correct value of these parameters, it is required to use a cutoff of $6\sigma$ and above.

In the future, we plan to apply this scheme to several other molecular systems, like gas-liquid nucleation in water and also to liquid-solid crystallization.

## ACKNOWLEDGMENTS

We thank Professor J.T. Hynes and Professor M. Klein for useful discussions. SA thanks the Indian Institute of Science for support. BB thanks DST-SERB for the grant of an India National Science Chair Professorship.

# Supplementary Material (SM)

## Contents

In this Supplementary Material part, we have presented further analyses with numerical results to support and supplement the main results of the text.

## S1. Degree of Supersaturation

The supersaturation (S) is computed as the ratio of the actual and equilibrium vapor pressure. The equilibrium vapor pressure of Argon at our thermodynamic condition is reported to be 0.43 bar.[1–3] We tune the pressure of the simulation box to achieve the desired supersaturation level $S$. **Table S1** reports the pressure and corresponding supersaturation used in our study.

**Table S1: Pressure and corresponding supersaturation invoked in our study**

| Supersaturation | Pressure of the system (bar) |
|---|---|
| | |
| 7.0 | 3.0 |
| 10.4 | 4.5 |
| 11.6 | 5.0 |
| 12.8 | 5.5 |

## S2. Order parameters against time

In the main manuscript, we have considered three order parameters, namely (i) the total number of liquid-like particles ($n_{liq}$) (ii) the second moment of the coordination number fluctuation ($\mu_2$), and (iii) the total number of liquid-like particles in the largest cluster ($n_{largest}$). Because of the high supersaturation in the present system, we can observe the gas-liquid nucleation even in the unbiased simulation. In this regard, we perform several unbiased simulations starting from the vapor phase and monitor the order parameters with time. In **Figure S1**, we show the time trajectories of these order parameters at S=12.8 by the dark red colored lines.

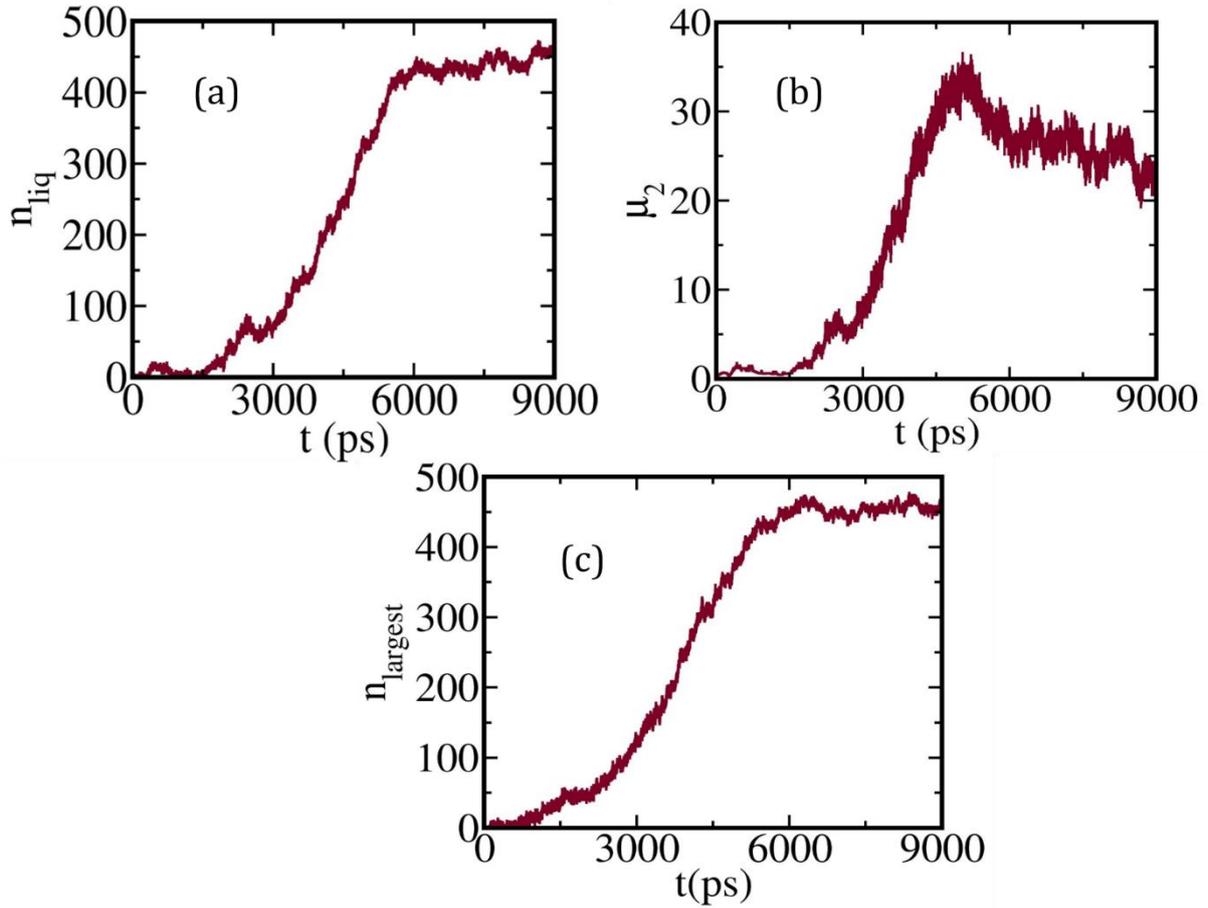

**Figure S1: Plot of the order parameters against time.** We perform an unbiased simulation at S=12.8 and monitor three order parameters with time. In (a), we plot the total number of liquid-like particles ($n_{liq}$) with time. (b) Plot of the second order of the coordination number fluctuation ($\mu_2$) against time. In (c), we plot the total number of liquid-like particles in the largest cluster ($n_{largest}$) with time.

## S3. Required parameters and the nucleation rate at S=7.0

As discussed in the main manuscript, we calculate the required parameters using Hynes formalism[4–6] to compute the nucleation rate at S=7.0. **Table S2** reports all the parameters required for the rate calculation.

**Table-S2: The required parameters for the calculation of nucleation rate at S=7.0 along the mutually orthogonal coordinates, X and Y. We have calculated the parameters by employing Hynes formalism.**

| Well frequency along X ($\omega_X^w$) | $\omega_X^w = (1.09 \pm 0.02)\,ps^{-1}$ |
|---|---|
| Well frequency along Y ($\omega_Y^w$) | $\omega_Y^w = (1.07 \pm 0.01)\,ps^{-1}$ |
| Barrier frequency along X ($\omega_X^b$) | $\omega_X^b = (3.0 \pm 0.02)\,ps^{-1}$ |
| Barrier frequency along Y ($\omega_Y^b$) | $\omega_Y^b = (2.8 \pm 0.01)\,ps^{-1}$ |
| Zero frequency Friction along X, $\zeta_{XX}(0)$ | $\zeta_{XX}(0) = (35.87 \pm 0.01)\,ps^{-1}$ |
| Friction along Y, $\zeta_{YY}(0)$ | $\zeta_{YY}(0) = (34.09 \pm 0.01)\,ps^{-1}$ |
| Free energy barrier ($\Delta G^\dagger$) | $\Delta G^\dagger = 9.63\,kJ/mol$ |

In **Table S3**, we provide the numerical values of the rate predicted by the Markovian and non-Markovian prescriptions at S=7.0.

**Table-S3: The numerical values of rate constants along the reaction coordinate *X* at S=7.0. Here, we have calculated the rate from different theoretical approaches. The appropriate parameters used here are given in Table S2.**

| Theoretical approach | Expressions | Numerical value |
|---|---|---|
| 1D transition state theory | $k_f^{TST} = \dfrac{\omega_X^w}{2\pi} \exp\left(-\dfrac{\Delta G^\dagger}{k_B T}\right)$ | $(101.09 \pm 0.01)\,ms^{-1}$ |
| 1D Kramers' theory | $k_f^{Kr} = \dfrac{1}{\omega_X^b}\left[\left(\dfrac{\zeta_{XX}^2}{4} + (\omega_X^b)^2\right)^{\frac{1}{2}} - \dfrac{\zeta_{XX}}{2}\right] k_f^{TST}$ | $(12.73 \pm 0.01)\,ms^{-1}$ |

| 1D Smoluchowski equation | $k_f^{SL} = \left(\dfrac{\omega_X^b}{\zeta_{XX}}\right) k_f^{TST}$ | $(12.86 \pm 0.01)\, ms^{-1}$ |
|---|---|---|
| 2D Transition state theory | $k_{2D}^{TST} = \left(\dfrac{\omega_Y^w}{\omega_Y^b}\right) k_f^{TST}$ | $(91.66 \pm 0.02)\, ms^{-1}$ |
| Kramers-Langer | $k_f^{KL} = \dfrac{\lambda_+}{2\pi}\left[\dfrac{\det \mathbf{E^w}}{|\det \mathbf{E^b}|}\right]^{1/2} \exp\left(-\dfrac{\Delta G^\dagger}{k_B T}\right)$ | $(11.58 \pm 0.01)\, ms^{-1}$ |
| Grote-Hynes theory with non-Markovian friction along both modes (with zero coupling friction) | $k_{GH} = \left(\dfrac{\lambda_X}{\omega_X^b}\right) k_{2d}^{TST}$ | $(20.94 \pm 0.02)\, ms^{-1}$ |
| Landauer and Swanson | $k_f^{LS} = \left(\dfrac{\omega_X^b}{\zeta_{XX}}\right) k_{2d}^{TST}$ | $(11.75 \pm 0.01)\, ms^{-1}$ |

## S4. Required parameters and the nucleation rate at S=10.4

In **Table S4**, we report all the parameters required for the rate calculation at S=10.4.

**Table-S4: The required parameters for the calculation of nucleation rate along the mutually orthogonal coordinates, X and Y. We have calculated the parameters by employing Hynes formalism.**

| Well frequency along X ($\omega_X^w$) | $\omega_X^w = (1.03 \pm 0.02)\, ps^{-1}$ |
|---|---|
| Well frequency along Y ($\omega_Y^w$) | $\omega_Y^w = (1.03 \pm 0.01)\, ps^{-1}$ |
| Barrier frequency along X ($\omega_X^b$) | $\omega_X^b = (2.27 \pm 0.02)\, ps^{-1}$ |
| Barrier frequency along Y ($\omega_Y^b$) | $\omega_Y^b = (1.64 \pm 0.01)\, ps^{-1}$ |

| **Zero frequency Friction along X**, $\zeta_{XX}(0)$ | $\zeta_{XX}(0) = (30.86 \pm 0.01) ps^{-1}$ |
|---|---|
| **Friction along Y**, $\zeta_{YY}(0)$ | $\zeta_{YY}(0) = (31.08 \pm 0.01) ps^{-1}$ |
| **Free energy barrier** ($\Delta G^\dagger$) | $\Delta G^\dagger = 5.59 kJ/mol$ |

Now, we have all the required parameters for the rate calculation. It's time to calculate the rate constants according to the different theoretical prescriptions by using these parameters as input. In **Table S5**, we provide the numerical values of the rate predicted by the Markovian and non-Markovian prescriptions at S=10.4.

**Table-S5: The numerical values of rate constants along the reaction coordinate $X$ at S=10.4. Here, we have calculated the rate from different theoretical approaches. The appropriate parameters used here are given in Table S4.**

| Theoretical approach | Expressions | Numerical value |
|---|---|---|
| **1D transition state theory** | $k_f^{TST} = \dfrac{\omega_X^w}{2\pi} \exp\left(-\dfrac{\Delta G^\dagger}{k_B T}\right)$ | $(39.06 \pm 1.0) \mu s^{-1}$ |
| **1D Kramers' theory** | $k_f^{Kr} = \dfrac{1}{\omega_X^b}\left[\left(\dfrac{\zeta_{XX}^2}{4} + (\omega_X^b)^2\right)^{\frac{1}{2}} - \dfrac{\zeta_{XX}}{2}\right] k_f^{TST}$ | $(5.83 \pm 0.1) \mu s^{-1}$ |
| **1D Smoluchowski equation** | $k_f^{SL} = \left(\dfrac{\omega_X^b}{\zeta_{XX}}\right) k_f^{TST}$ | $(5.96 \pm 0.1) \mu s^{-1}$ |
| **2D Transition state theory** | $k_{2D}^{TST} = \left(\dfrac{\omega_Y^w}{\omega_Y^b}\right) k_f^{TST}$ | $(24.53 \pm 0.11) \mu s^{-1}$ |

| Kramers-Langer | $k_f^{KL} = \frac{\lambda_+}{2\pi} \left[ \frac{\det \mathbf{E^w}}{|\det \mathbf{E^b}|} \right]^{1/2} \exp\left(-\frac{\Delta G^\dagger}{k_B T}\right)$ | $(3.64 \pm 0.2)\,\mu s^{-1}$ |
|---|---|---|
| Grote-Hynes theory with non-Markovian friction along both modes (with zero coupling friction) | $k_{GH} = \left(\frac{\lambda_X}{\omega_X^b}\right) k_{2d}^{TST}$ | $(10.10 \pm 0.2)\,\mu s^{-1}$ |
| Landauer and Swanson | $k_f^{LS} = \left(\frac{\omega_X^b}{\zeta_{XX}}\right) k_{2d}^{TST}$ | $(3.65 \pm 0.1)\,\mu s^{-1}$ |

## S5. Required parameters and the nucleation rate at S=12.8

**Table S6** reports all the necessary parameters for the rate calculation at S=12.8.

**Table-S6:** The required parameters for the calculation of nucleation rate along the mutually orthogonal coordinates, X and Y, at S=12.8. We have calculated the parameters by employing Hynes formalism.

| Well frequency along X ($\omega_X^w$) | $\omega_X^w = (1.0 \pm 0.02)\,ps^{-1}$ |
|---|---|
| Well frequency along Y ($\omega_Y^w$) | $\omega_Y^w = (0.99 \pm 0.01)\,ps^{-1}$ |
| Barrier frequency along X ($\omega_X^b$) | $\omega_X^b = (1.58 \pm 0.02)\,ps^{-1}$ |
| Barrier frequency along Y ($\omega_Y^b$) | $\omega_Y^b = (1.45 \pm 0.01)\,ps^{-1}$ |
| Zero frequency Friction along X, $\zeta_{XX}(0)$ | $\zeta_{XX}(0) = (15.48 \pm 0.01)\,ps^{-1}$ |
| Friction along Y, $\zeta_{YY}(0)$ | $\zeta_{YY}(0) = (15.62 \pm 0.01)\,ps^{-1}$ |
| Free energy barrier ($\Delta G^\dagger$) | $\Delta G^\dagger = 3.71\,kJ/mol$ |

In **Table S7**, we provide the numerical values of the rate predicted by the Markovian and non-Markovian prescriptions at S=12.8.

**Table-S7: The numerical values of rate constants along the reaction coordinate *X* at S=12.8. Here, we have calculated the rate from different theoretical approaches. The appropriate parameters used here are given in Table S6.**

| Theoretical approach | Expressions | Numerical value |
|---|---|---|
| **1D transition state theory** | $k_f^{TST} = \dfrac{\omega_X^w}{2\pi} \exp\left(-\dfrac{\Delta G^\dagger}{k_B T}\right)$ | $(631.48 \pm 1.0)\,\mu s^{-1}$ |
| **1D Kramers' theory** | $k_f^{Kr} = \dfrac{1}{\omega_X^b}\left[\left(\dfrac{\zeta_{XX}^2}{4} + (\omega_X^b)^2\right)^{\frac{1}{2}} - \dfrac{\zeta_{XX}}{2}\right] k_f^{TST}$ | $(63.79 \pm 0.1)\,\mu s^{-1}$ |
| **1D Smoluchowski equation** | $k_f^{SL} = \left(\dfrac{\omega_X^b}{\zeta_{XX}}\right) k_f^{TST}$ | $(64.45 \pm 0.1)\,\mu s^{-1}$ |
| **2D Transition state theory** | $k_{2D}^{TST} = \left(\dfrac{\omega_Y^w}{\omega_Y^b}\right) k_f^{TST}$ | $(428.78 \pm 0.11)\,\mu s^{-1}$ |
| **Kramers-Langer** | $k_f^{KL} = \dfrac{\lambda_+}{2\pi}\left[\dfrac{\det \mathbf{E^w}}{|\det \mathbf{E^b}|}\right]^{1/2} \exp\left(-\dfrac{\Delta G^\dagger}{k_B T}\right)$ | $(43.14 \pm 0.31)\,\mu s^{-1}$ |
| **Grote-Hynes theory with non-Markovian friction along both modes (with zero coupling friction)** | $k_{GH} = \left(\dfrac{\lambda_X}{\omega_X^b}\right) k_{2d}^{TST}$ | $(100.95 \pm 0.2)\,\mu s^{-1}$ |
| **Landauer and Swanson** | $k_f^{LS} = \left(\dfrac{\omega_X^b}{\zeta_{XX}}\right) k_{2d}^{TST}$ | $(43.76 \pm 0.1)\,\mu s^{-1}$ |

## S6. Finite size effects

To investigate the impact of finite size on gas-liquid nucleation, we systematically vary the system size while maintaining a constant supersaturation, i.e., S=11.6.[7] Specifically, we consider system sizes of 256, 512, and 1024, conducting unbiased simulations to construct the free energy surface based on two order parameters: the liquidness of the system ($n_{liq}$) and the second moment of the coordination number fluctuation ($\mu_2$).

We calculate the energy barrier from the resulting free energy surface for various system sizes and present the findings in **Figure S2**. Notably, **Figure S2** demonstrates that the barrier height, a critical factor influencing nucleation rate, exhibits high sensitivity to the system's size. This sensitivity contributes to the significant disparity observed in nucleation rates between experimental and simulation results.

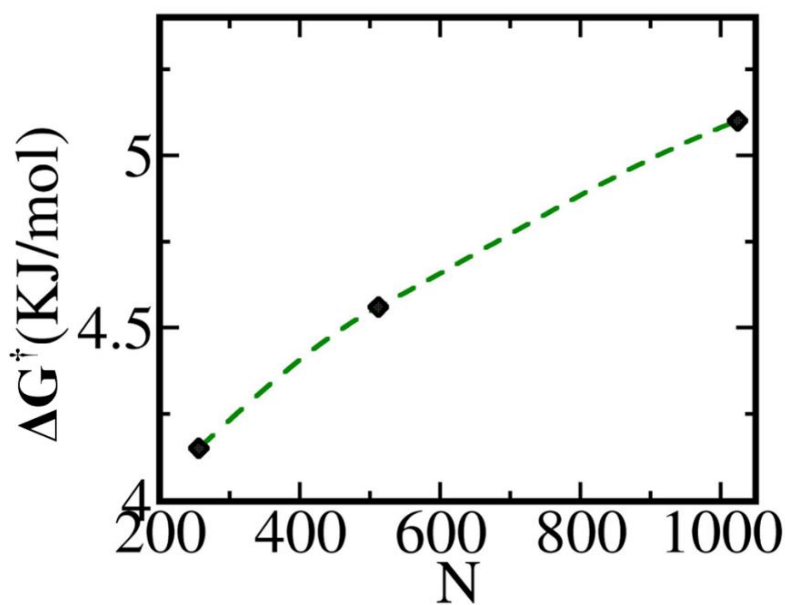

**Figure S2: Plot of free energy against the system size. Clearly, barrier height increases with the size of the system. Here, the green dashed lines joining the data points are provided as a guide to the eyes.**

# References for SM